\documentclass[amsmath,notitlepage,amssymb,aps,showkeys,floatfix,prd,a4paper,
  onecolumn,nofootinbib]{revtex4-2}
\usepackage{times,amsbsy,amsfonts,graphicx,float}
\usepackage{color,morefloats,rotating,srcltx,slashed}
\usepackage{multirow,bm,verbatim,tabularx,bbding,threeparttable}
\definecolor{dblue}{rgb}{0.00,0.00,0.75}
\usepackage[colorlinks,urlcolor=dblue,linkcolor=dblue,citecolor=dblue]{hyperref} 
\usepackage{setspace}
\allowdisplaybreaks[4]

\begin{document} 

\title{Role of $f_0(980)$ and $a_0(980)$ in the $B^- \to \pi ^- K^+ K^- $ and $B^- \to \pi ^- K^0 \bar K^0 $ reactions} 

\author{Luciano M. Abreu$^{1,2}$}     \email{luciano.abreu@ufba.br}
\author{Natsumi Ikeno$^{3}$}     \email{ikeno@tottori-u.ac.jp}
\author{Eulogio Oset$^2$}             \email{oset@ific.uv.es}
		
\affiliation{$^1$Instituto de F\'{\i}sica, Universidade Federal da Bahia, Campus Universit\'{a}rio de Ondina, 40170-115 Bahia, Brazil\\
                 $^2$Departamento de F\'{\i}sica Te\'orica and IFIC, Centro Mixto Universidad de Valencia-CSIC, \\
	                 Institutos de Investigaci\'on de Paterna, Aptdo.~22085, 46071 Valencia, Spain\\           
	         $^3$Department of Agricultural, Life and Environmental Sciences,
Tottori University, Tottori 680-8551, Japan
}

\begin{abstract}

In this work we study the role of the $f_0(980)$ and $a_0(980)$ resonances in the low $ K ^{+} K^{-} $ and $K^0 \bar K^0 $ invariant-mass region of the $B^- \to \pi ^- K^+ K^- $ and $B^- \to \pi ^- K^0 \bar K^0 $ reactions. The amplitudes are calculated by using the chiral unitary $\rm SU(3)$ formalism, in which these two resonances are dynamically generated from the unitary pseudocalar-pseudoscalar coupled-channel approach. The amplitudes are then used as input in the evaluation of the mass distributions with respect to the  $  K^{+}K^{-} $ and  $ K^{0}\bar K^{0} $ invariant-masses, where the contributions coming from the $I=0$ and $I=1$ components are explicitly assessed. Furthermore, the contribution of the $ K^{\ast }(892)^0 K^- $ production and its influence on the $ \pi^{-} K^+ $ and $ K^{+} K^- $ systems are also evaluated, showing that there is no significant strength for small $ K^{+} K^- $ invariant mass. Lastly, the final distributions of $ M_{\rm inv}^2( K^{\pm}K^{\mp} ) $ for the  $B^{\mp} \to  \pi ^{\mp} K^{\pm}K^{\mp}  $ reactions are estimated and compared with the LHCb data. Our results indicate that the $I=0$ component tied to the $f_0(980)$ excitation generates the dominant contribution in the range of low $ K ^{+} K^{-} $ invariant-mass. 

\end{abstract}

\date{\today}

\maketitle

\section{Introduction}                  

%

Hadronic and charmless three-body decays of $B$ mesons have become a prominent testing ground for studying the hadron dynamics and, in a more profound sense, the validity limits of the Standard Model. According to it, these decays are suppressed, and therefore unexpected enhancements in the branching fractions might indicate a physics beyond the Standard Model. That being so, one can naturally wonder about the direct CP violation in these decays, since they can present large CP asymmetries coming from the interference of tree and loop diagrams. In this scenario, beyond Standard Model particles in principle might contribute in these loop diagrams~(for a more detailed discussion see Ref.~\cite{Belle:2022bpm}). As emblematic examples, previous experimental investigations on the CP violation in $B^{+} \to \pi ^{+} K^+ K^- $ decay have been performed by Babar and Belle collaborations~\cite{BaBar:2007itz,Belle:2017cxf}.  Babar~\cite{BaBar:2007itz} reported the first measurement of the branching fraction of this decay, $\mathcal{B}=[5.0 \pm 0.5 (stat) \pm 0.5 (syst) ]\times 10^{-6}$, and found CP asymmetry consistent with zero, while Belle in~\cite{Belle:2017cxf} observed a strong evidence of a large direct CP asymmetry in the low $ K ^{+} K^{-} $ invariant-mass region.

Recently, in a series of works the LHCb collaboration has found the direct CP violation in charmless three-body decays of $B$ mesons~\cite{LHCb:2013ptu,LHCb:2013lcl,LHCb:2014mir,LHCb:2019xmb,LHCb:2019jta,LHCb:2019sus}. In particular, in Ref.~\cite{LHCb:2019xmb} the first amplitude analysis of the $B^{\pm} \to \pi ^{\pm} K^+ K^- $ decay has been performed, by considering contributions of the resonances $ K^{\ast }(892)^0 $ and $ K^{\ast }_0(1430)^0 $ plus a nonresonant contribution in the final state $ \pi ^{\pm} K^{\mp} $ and the resonances $ \phi (1020) $, $ f_2 (1270) $ and $ \rho (1430) $  plus a component from $S$-wave $\pi \pi \leftrightarrow K K $ rescattering in the $ K ^{\pm} K^{\mp} $ system. The total decay amplitude is modelled  via the the isobar model, with the resonant structures being associated to a relativistic Breit-Wigner lineshape function; also, a single-pole form factor accounts for the nonresonant amplitude.  The data are found to be well described by the coherent sum of these five resonant structures plus a nonresonant contribution and $\pi \pi \leftrightarrow K K $ rescattering. Interestingly, this last contribution has a sizable fit fraction and acquires the largest CP asymmetry in the low $ K ^{\pm} K^{\mp} $ invariant-mass region. Specifically, it has been encoded in a $S$-wave $\pi \pi \leftrightarrow K K $ transition amplitude with isospin $I=0$ and total angular momentum $J=0$, and given by the off-diagonal term in the $S$-matrix for the $\pi \pi$ and $ K K $  coupled channel.

On theoretical grounds, the $B^{\pm} \to \pi ^{\pm} K^+ K^- $ decays have also gained attention in the last years~\cite{Cheng:2007si,Cheng:2013dua,Cheng:2016shb,Cheng:2020ipp,Wang:2020plx,Wang:2020nel,Shi:2021ste}. Refs.~\cite{Cheng:2007si,Cheng:2013dua,Cheng:2016shb,Cheng:2020ipp} have employed the factorization approach. Looking specifically at the appproach of Ref.~\cite{Cheng:2020ipp}, the amplitude is decomposed as the coherent sum of resonant contributions together with the nonresonant background, as well. The resonant amplitudes are also related to quasi-two-body decay processes and described by the relativistic Breit-Wigner lineshape model,  in which contributions from $ K^{\ast }(892)^0, f_0(980), \phi (1020), f_2 (1270) , K^{\ast }_0(1430)^0 $ and $ \rho (1430) $ were considered; and the nonresonant contribution is parameterized in terms of form factors based on heavy meson chiral perturbation theory. Besides, final-state rescattering of $S$-wave $\pi^+ \pi^- \leftrightarrow K^+ K^- $ is also taken into account. The calculated branching fractions of resonant and nonresonant contributions are found to be in some case in contrast with respect to the LHCb results in~\cite{LHCb:2019xmb}. But we should remark that in the approach of~\cite{Cheng:2020ipp} the $f_0(980) $, not considered in the fit model of LHCb, has a nonnegligible contribution with respect to the other resonant structures, with branching fraction $\mathcal{B}=(0.19\pm 0.03)\times 10^{-6}$. 

It is also worth mentioning the other attempts of description of the LHCb analysis. 
For instance, Ref.~\cite{Wang:2020plx} has analyzed the quasi-two-body decays $B^{+} \to \pi^{+} \rho (770,1450)^0 \to \pi^{+}  K^+ K^- $ in the perturbative QCD approach, which contribute about $5 \% $ of the total branching fraction, 
much less than the $(30.7 \pm 1.2 \pm 0.9 ) \% $ from LHCb~\cite{LHCb:2019xmb} for the 
$\rho (1450)^0$ contribution. It has been suggested that  the absence of the $\rho (770)^0 \to K^+ K^- $ in the decay amplitude of three-body $B$ decays could probably result in a larger proportion for the resonance $\rho (1450)^0$ in the experimental amplitude analysis.

In a different perspective, the work~\cite{Shi:2021ste} has studied the resonant contribution to the decay amplitude of $B^{-} \to \pi ^{-} K^+ K^- $  dominated by the $ K^{\ast }(892)^0, f_0(980), \phi (1020), f_2 (1270) , K^{\ast }_0(1430)^0 $ resonances, where the quasi two-body decays have been calculated within the light-cone sum rule approach utilizing the leading twist $B$ meson light-cone distribution amplitudes. Some branching fraction have been found consistent with experiment, while the others are smaller than the measured values. The authors argue that one possible reason for this discrepancy might be the uncertainties of the strong couplings between the corresponding resonance with pseudoscalar mesons. In particular, the branching fraction for the $\rho (1450)^0$ contribution was about one order smaller than that of LHCb. Notably, the result for the $f_0(980)$ resonance ($\mathcal{B}=(0.12\pm 0.04)10^{-6}$)) is consistent with that from Ref.~\cite{Cheng:2020ipp}, still waiting for future experimental tests. 

From the discussion above one can conclude that both experimental and theoretical amplitude analyses of $B \to \pi  K K $  decays are involved, remaining as a matter of debate. The interferences relating the resonant structures as well as the nonresonant amplitude complicate the evaluation and identification of the nonresonant and resonant contributions.

In this sense, we intend to contribute on this subject with a distinct viewpoint of the preceding works. Benefiting from the previous investigations~\cite{Cheng:2020ipp,Shi:2021ste} that pointed out the possible relevance of the $f_0(980)$ contribution, in the present study we analyze the role of the $f_0(980)$ and $a_0(980)$ resonances in the low $ K ^{+} K^{-} $ invariant-mass region of the $B^- \to \pi ^- K^+ K^- $ and $B^- \to \pi ^- K^0 \bar K^0 $ reactions. The amplitudes are calculated by using the chiral unitary $\rm SU(3)$ formalism, in which these two resonances are dynamically generated from the unitary pseudocalar-pseudoscalar coupled-channel approach. Then, we compute the mass distributions with respect to the  $  K^{+}K^{-} $ and  $ K^{0}\bar K^{0} $ invariant-masses, where the contributions coming from the $I=0$ and $I=1$ components are explicitly assessed. Additionally, we also calculate the contribution of the $ K^{\ast }(892)^0 K^- $ production on the  $ \pi^-  K^+$ and  $ K^{+} K^- $  systems. Finally, the distributions of $ M_{\rm inv}^2( K^{\pm}K^{\mp} ) $ for the  $B^{\mp} \to  \pi ^{\mp} K^{\pm}K^{\mp}  $ reaction are estimated and compared with the LHCb data in~\cite{LHCb:2019xmb}. Our approach has a strong similarity to the one used in related works on the $D_s^+ \to \pi ^+ K^+ K^- $ decay in \cite{Wang:2021ews} and $D_s^+ \to \pi ^0 K^+ K_S^0 $ decay in \cite{Zhu:2022guw}, where the $f_0(980)$ and $a_0(980)$ resonances are generated in the same way and a good description of the data is obtained.

\section{Formalism}\label{sec:2}           

\subsection{Transition matrix and mass distributions: external emission mechanism}\label{sec:2.1}  

\begin{figure}[tbp]   
  \centering
  \includegraphics[width=8cm]{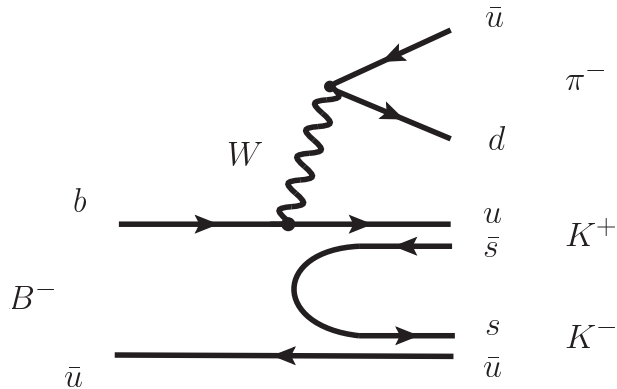}
  \caption{Mechanism at the quark level for the production of the $B^- \to \pi^- K^+ K^- $ reaction.
              }
  \label{fig1}
\end{figure}

We start by considering the $B^- \to \pi ^- K^+ K^- $ reaction. The mechanism at the quark level for the production of the final state considered here is the Cabibbo-suppressed external emission diagram, depicted in Fig.~\ref{fig1}. 
The hadronization is performed including a $\bar{q} q$ pair with the quantum numbers of the vacuum. Accordingly, denoting $ \bar{q} q  \equiv \sum_{i} \bar{q}_i q_i$ ($i=\left\{u,d,s\right\}$), we obtain two pseudoscalar mesons as follows,
\begin{eqnarray}
u \bar{u} & \to & \sum_{i} u \bar{q}_i q_i  \bar{u} \to (PP)_{11}, 
\label{eq1}
\end{eqnarray}
where $ P$ is the  $ q \bar{q} $ matrix in $ \text{SU}(3) $ flavor space written in terms of pseudoscalar mesons:
\begin{equation}
P = \left(
           \begin{array}{ccc}
             \frac{1}{\sqrt{2}}\pi^0 + \frac{1}{\sqrt{3}}\eta + \frac{1}{\sqrt{6}}\eta' & \pi^+ & K^+  \\
             \pi^- & -\frac{1}{\sqrt{2}}\pi^0 + \frac{1}{\sqrt{3}}\eta + \frac{1}{\sqrt{6}}\eta' & K^0\\
            K^- & \bar{K}^0 & -\frac{1}{\sqrt{3}}\eta + \sqrt{\frac{2}{3}}\eta'  \\
          \end{array}
         \right).
  \label{eq2}
\end{equation}
where the standard $\eta - \eta^{\prime}$ mixing of Ref.~\cite{Bramon:1994cb} has been considered. 

Ignoring the terms involving the $\eta^{\prime}$, then we can obtain the following combination 
\begin{eqnarray}
u \bar{u} & \to & (PP)_{11} = C  \left( \frac{1}{2}\pi^0\pi^0 + \pi^+\pi^- + \frac{1}{3}\eta \eta +  \frac{2}{\sqrt{6}}\pi^0\eta + K^+ K^-  \right) ,
\label{eq3}
\end{eqnarray}
where $C$ is a constant to be fixed. 

In the combination given by Eq.~(\ref{eq3}) there are contributions with isospin $I=0$ and $I=1$. In this sense, it is convenient to write the $PP$ states in terms of the isospin states: 

\begin{itemize}

\item $I=0$
\begin{eqnarray}
&&|\pi\pi, I=0 \rangle = \frac{(-1)}{\sqrt6}(\pi^{+}\pi^{-}+\pi^{-}\pi^{+}+\pi^{0}\pi^{0}),   \nonumber\\
&&|K\bar K, I=0\rangle = \frac{(-1)}{\sqrt2}(K^{+}K^{-}+K^{0}\bar K^{0}), \nonumber\\
&&| \eta\eta \rangle \to \frac{1}{\sqrt2}| \eta\eta \rangle,
\label{eq4} 
\end{eqnarray}
where the isospin multiplets are defined as $(K^{+}, K^{0})$, $(\bar K^{0}, -K^{-})$, $(-\pi^+,\pi^0,\pi^-)$; and the $\frac{1}{\sqrt2}$ factor is used due to the unitary normalization adopted for identical particles in the counting of states in the intermediate loops. 

\item $I=1$
\begin{eqnarray}
&&|K\bar K, I=1 , I_3 = 0 \rangle = \frac{(-1)}{\sqrt2}(K^{+}K^{-}-K^{0}\bar K^{0}), \nonumber\\
&&| \pi \eta \rangle \equiv | \pi \eta ; I=1, I_3=0 \rangle = |  \pi^0 \eta \rangle.
\label{eq5} 
\end{eqnarray}

\end{itemize}

\begin{figure}[tbp]   
  \centering
  \includegraphics[width=6.5cm]{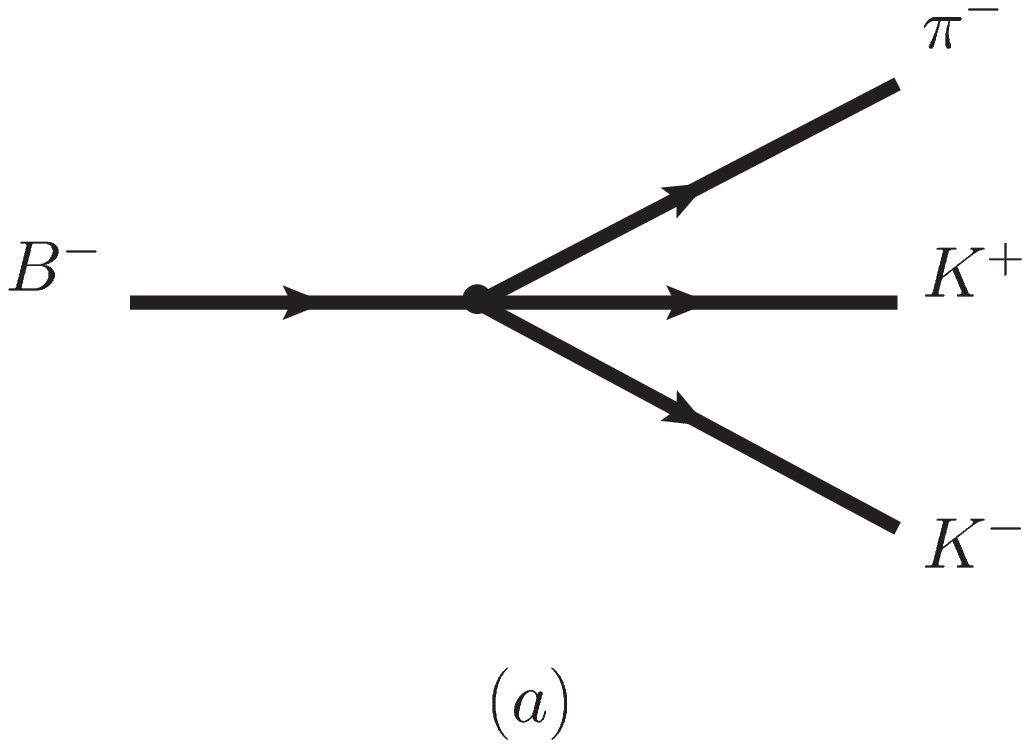}
  \includegraphics[width=8cm]{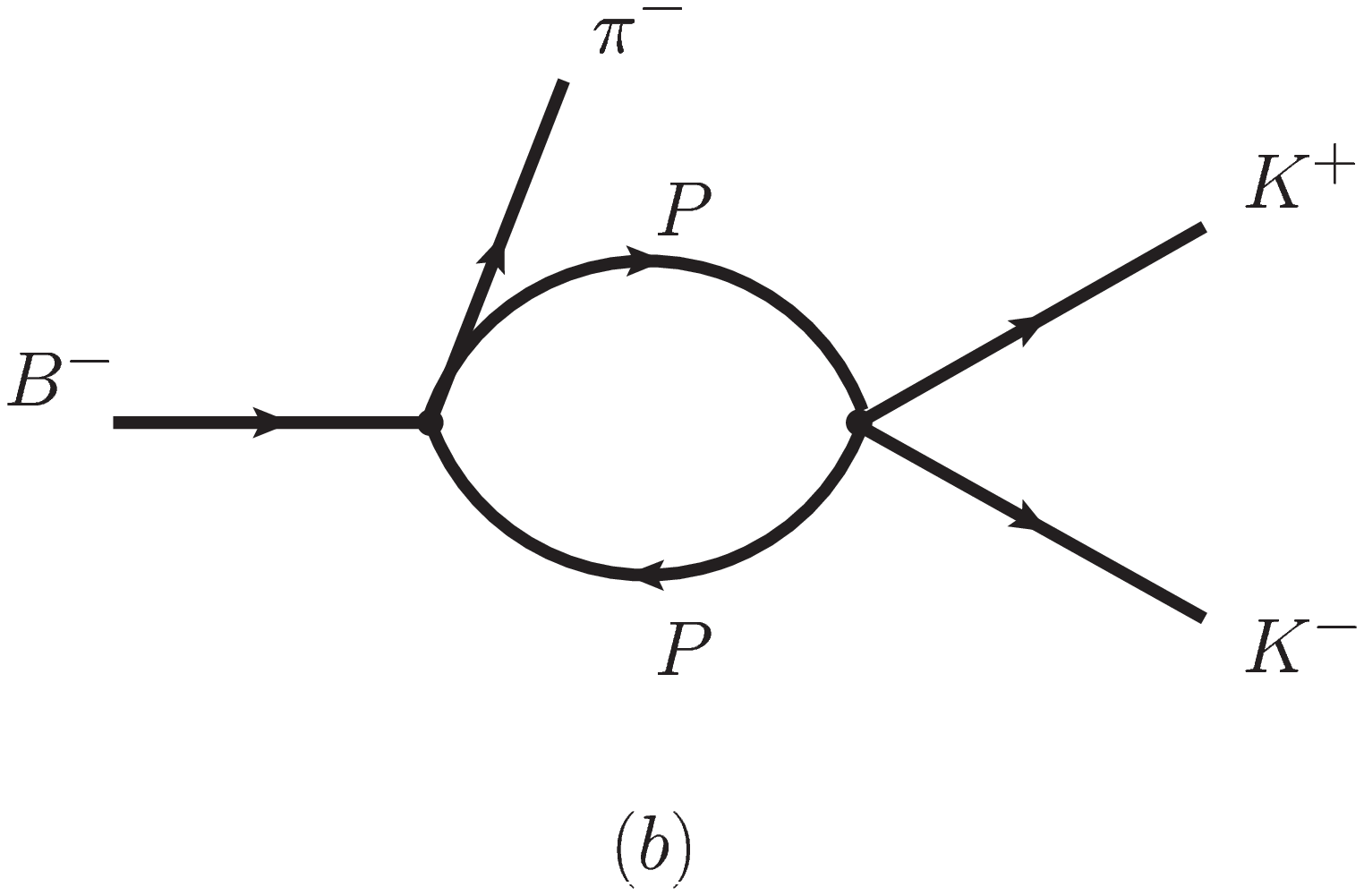}
   \includegraphics[width=8cm]{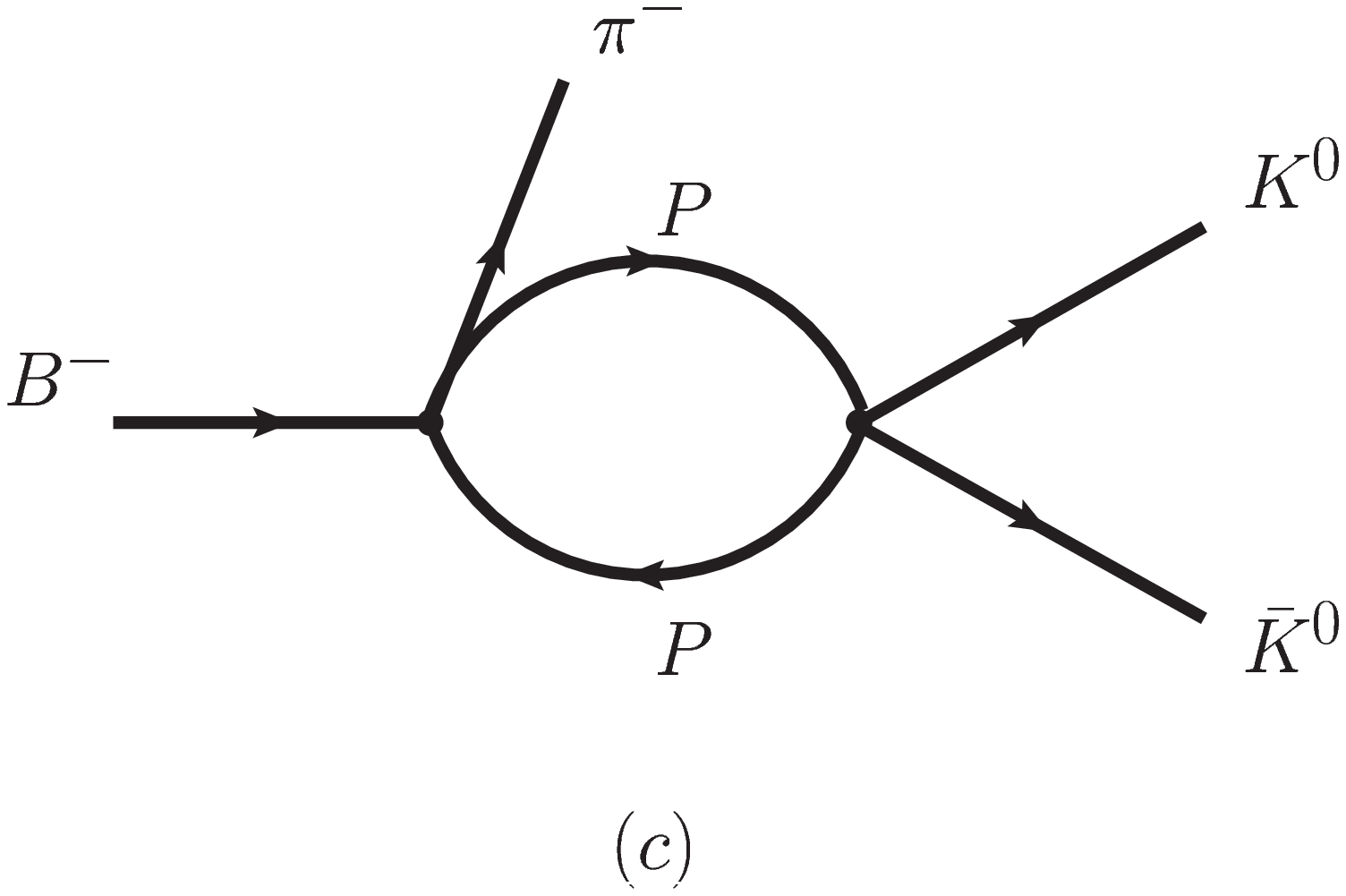}
   \caption{Mechanisms for the production of the $B^- \to  \pi ^- K^+ K^- $[(a) and (b)] and $B^- \to  \pi ^- K^{0}\bar K^{0}$ [(c)] reactions.
              }
  \label{fig2}
\end{figure}

Then, the structure in Eq.~(\ref{eq1}) can yield channels with $I=0$ and $I=1$; however, at tree level only the reaction with content $ K^{+}K^{-} $ in the final state is generated. The $ K^{0}\bar K^{0} $ channel will be produced through rescattering. These production mechanisms are shown in Fig.~\ref{fig2}. As a consequence, the effects of the states $f_0(980)$ and $a_0(980)$ will be present in the rescattering contributions, since the $PP$ pairs should interact and produce these scalar resonances.

Thus, the analytical expressions of the transition matrices associated to the mechanisms depicted in Fig.~\ref{fig2} can be written as 
\begin{eqnarray}
t_{K^{+}K^{-}}(M_{\rm inv}) &  = & C + C  \left(\frac{-1}{\sqrt2}\right) \left[ W_{ \pi \pi} \, G_{ \pi \pi}(M_{\rm inv})\, T_{ \pi \pi ,K\bar K}^{\,(I=0)}  (M_{\rm inv})
+ W_{  \eta \eta} \, G_{  \eta \eta}(M_{\rm inv}) \, T_{  \eta \eta ,K\bar K}^{\,(I=0)} (M_{\rm inv})
\right. \nonumber \\
& & + \left. W_{   K^{+}K^{-} }^{(I=0)} \, G_{   K \bar K }(M_{\rm inv})\, T_{ K\bar K ,K\bar K}^{\,(I=0)}  (M_{\rm inv})
\right. \nonumber \\
& &  \left. +  W_{   K^{+}K^{-} }^{(I=1)}  \, G_{   K \bar K }(M_{\rm inv})\, T_{   K\bar K ,K\bar K}^{\,(I=1)}  (M_{\rm inv})
+ W_{   \pi \eta } \, G_{   \pi \eta  }(M_{\rm inv})\, T_{   \pi \eta , K\bar K}^{\,(I=1)}(M_{\rm inv}) \right] ,    
\nonumber \\
t_{K^{0}\bar K^{0}}(M_{\rm inv}) &  = &  C  \left[ \left(\frac{-1}{\sqrt2}\right) W_{ \pi \pi} \, G_{ \pi \pi}(M_{\rm inv})\, T_{ \pi \pi ,K\bar K}^{\,(I=0)}  (M_{\rm inv})
+ \left(\frac{-1}{\sqrt2}\right)  W_{  \eta \eta} \, G_{  \eta \eta}(M_{\rm inv}) \, T_{  \eta \eta ,K\bar K}^{\,(I=0)} (M_{\rm inv})
\right. \nonumber \\
& & + \left. \left(\frac{-1}{\sqrt2}\right)  W_{   K^{+}K^{-} }^{(I=0)} \, G_{   K \bar K }(M_{\rm inv})\, T_{ K\bar K  ,K\bar K}^{\,(I=0)}  (M_{\rm inv})
\right. \nonumber \\
& & + \left. \left(\frac{1}{\sqrt2}\right)   W_{   K^{+}K^{-} }^{(I=1)}  \, G_{   K \bar K }(M_{\rm inv})\, T_{ K\bar K ,K\bar K}^{\,(I=1)}  (M_{\rm inv})
+ \left(\frac{1}{\sqrt2}\right) W_{   \pi \eta } \, G_{   \pi \eta  }(M_{\rm inv})\, T_{   \pi \eta   ,K\bar K}^{\,(I=1)}(M_{\rm inv}) \right] ,               
\label{tranmatrix}
\end{eqnarray}
where $ W_i 's$ are the weights calculated from the relationship between each combination in Eq.~(\ref{eq3}) and the corresponding isospin state in Eqs.~(\ref{eq4}) and~(\ref{eq5}), and are summarized in Table~\ref{weights};  $ G_i (M_{\rm inv}) $ is the loop function of the two intermediate pseudoscalar mesons; $ M_{\rm inv} $ is the invariant mass of the $ K \bar K $ system; finally, $ T_{i,j} $ represents the elements of the  unitarized transition matrix between $i = \pi \pi ,  K^{+}K^{-} (I=0) , \eta \eta ,  K^{+}K^{-} (I=1) , \pi \eta $ and $ j= K^{+}K^{-}$ or $ K^{0}\bar K^{0} $ states, obtained in Refs.~\cite{Oller:1997ti,Liang:2014tia,Xie:2014tma}
 from 
\begin{eqnarray}
      T=\left[1-VG\right]^{-1}V~,   \label{eq6}
\end{eqnarray} 
with $ V $ here denoting the interaction potential matrix. We use the same $ G $ and $ T $ matrices as in Refs.~\cite{Liang:2014tia,Xie:2014tma}, and accordingly the resonances $f_0(980)$ and $a_0(980)$ are produced by employing the unitarized $T$ matrix approach. The loop function $G$ appearing in Eqs.~(\ref{tranmatrix}) and~(\ref{eq6}) is regularized with cut-off regularization~\cite{Oller:1997ti}, with the value of cut-off used being $600$ MeV.

\begin{table}[!]
\centering
\caption{Summary of the weights  $ W_i \, (i = \pi \pi ,  K^{+}K^{-} (I=0) , \eta \eta ,  K^{+}K^{-} (I=1) , \pi \eta )$ calculated from the relationship between each combination in Eq.~(\ref{eq3}) and the corresponding isospin state in Eqs.~(\ref{eq4}) and~(\ref{eq5}). }
\label{weights}
\setlength{\tabcolsep}{8pt}
\setstretch{1.2}
\begin{tabular}{lccccc}
\hline 

\hline 
       &  $ \pi \pi $  & $ K^{+}K^{-} (I=0)$  & $\eta \eta $  & $ K^{+}K^{-} (I=1) $ & $ \pi \eta  $\\
\hline
$ W_i $ & $-\frac{1}{2}\sqrt{\frac{3}{2}}$  & $ -\frac{1}{\sqrt2} $ & $ \frac{\sqrt{2}}{3}$ & $-\frac{1}{\sqrt2}$ & $\sqrt{\frac{2}{3}}$ \\
\hline
\end{tabular}
\end{table}

Hence, one can remark two important differences between the transition elements in Eq.~(\ref{tranmatrix}): (i) the tree-level contribution is only present in $ t_{K^{+}K^{-}} $ production; and (ii) the interference among the contributions with $I=0$ and $I=1$ for $ t_{K^{+}K^{-}} $ is constructive, whereas in the case $t_{K^{0}\bar K^{0}}$ it is destructive.

The amplitudes in Eq.~(\ref{tranmatrix}) will be used in the standard expression of the mass distribution, 
\begin{eqnarray}
  \frac{d\Gamma_j}{dM_{\rm inv}}=\frac{1}{(2\pi)^3}\frac{1}{4 m^2_{B}} p_{\pi^-} \tilde {p}_{K_j} | t_{j} |^2,
      \label{eq7}
\end{eqnarray}
where  $ j= K^{+}K^{-}$ or $ K^{0}\bar K^{0} $, and
\begin{eqnarray}
    p_{\pi^-}&= & \frac{\lambda^{1/2}(m_{B^-}^2,m^2_{\pi^-},M_{\rm inv}^2)}{2m_{B^-}},  \label{eq8}    \\
    \tilde{p}_{K_j} & = & \frac{\lambda^{1/2}(M_{\rm inv}^2,m_{K_j}^2,m_{K_j}^2)}{2M_{\rm inv}}, 
    \label{eq9}
\end{eqnarray}
with $ \lambda(a,b,c) $ being the K\"all\'en function. We then get the two mass distributions that will allow us to evaluate the effects of the resonances $f_0(980)$ and $a_0(980)$.

%
%
%

\subsection{Contribution of the channel $ K^{\ast }(892)^0 K^- $}\label{sec:2.3}  

\begin{figure}[tbp]   
  \centering
  \includegraphics[width=8cm]{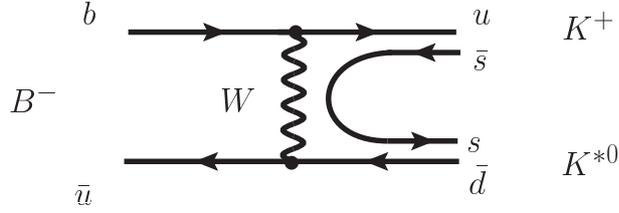}
  \caption{Mechanism at the quark level for the production of the $ B^- \to  K^{\ast }(892)^0 K^-  $ reaction from a $W$-exchange.
              }
  \label{fig3}
\end{figure}

One might ask about other possible mechanisms relevant in the present context. For example, at quark level a final state with $ K^-  K^{+} \pi^- $ might be produced from a $W$-exchange diagram, as depicted in Fig.~\ref{fig3}, via the $ K^{\ast }(892)^0 K^- $ production. 
However, Table I of Ref.~\cite{LHCb:2019xmb} shows that this reaction is suppressed with respect to the others. Besides, we must take into account that (i) the magnitude of the momentum of the $ K^- $ meson should be $p_{K^-} \simeq 2540  \,\rm MeV$; (ii) the $ K^{\ast 0} $ decays into $ K^{+} \pi^- $; (iii) the  $ K^+ $ meson produced from the decay has momentum of the order $p_{K^+} \simeq p_{K^-} [m_K/(m_{\pi}+m_{K})] \simeq 0.78 \, p_{K^-} \simeq 1980 \, \rm MeV$; (iv) $ M_{\rm inv} (K^{+}K^{-}) \simeq \sqrt{ ( E_{K^+}+E_{K^-} )^2 -  (p_{K^+}+p_{K^-} )^2 } \simeq 4592 \, \rm MeV$.  Hence, this mechanism presumably does not affect the  $ K^{+}K^{-} $ threshold energy. The same could be said if we produce a $ K_0^{\ast} (700) $ state that decays into $ K^{+} \pi^- $ in $S$-wave: its contribution should be very far away from the $K\bar K$ threshold.  

\begin{figure}[tbp]   
  \centering
  \includegraphics[width=8cm]{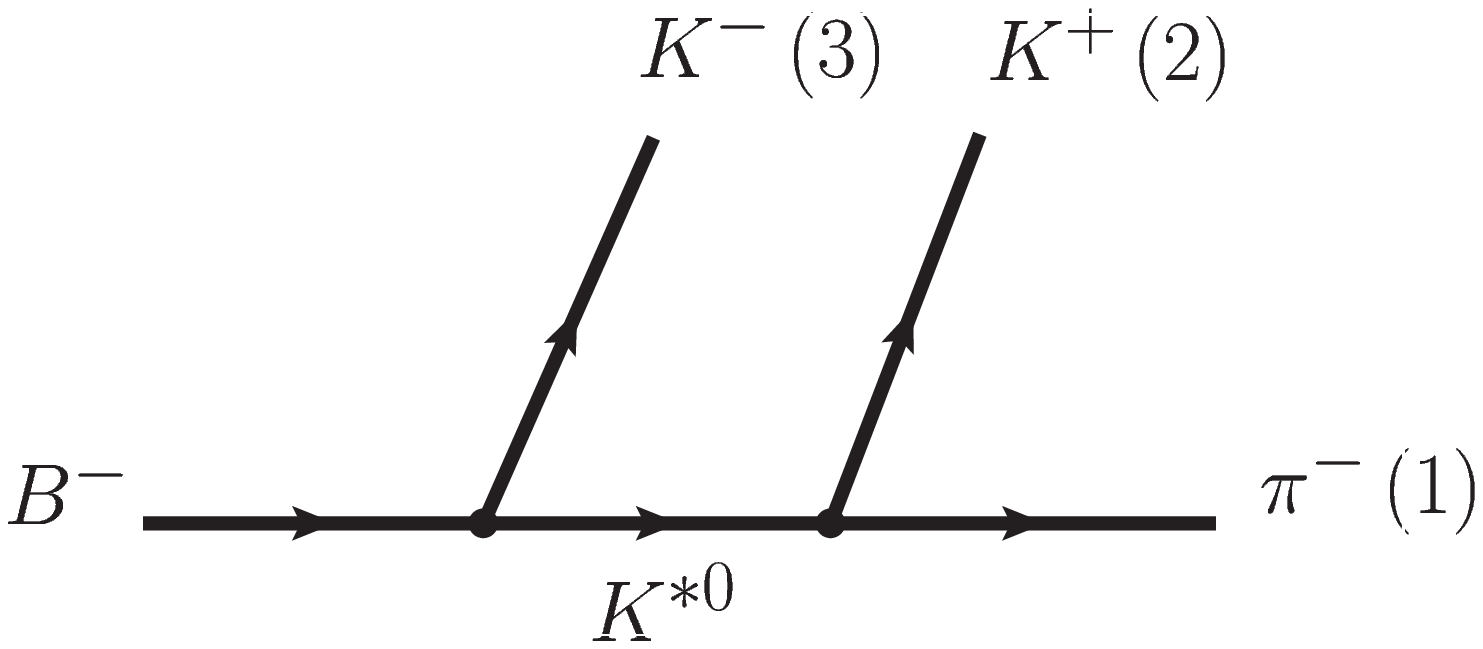}
  \caption{Mechanism for the production of the $ B^- \to  K^{\ast }(892)^0 K^- \to K^{+} \pi^-  K^- $  reaction.
              }
  \label{fig4}
\end{figure}

Notwithstanding, in the present approach we take into account the contribution of the $ K^{\ast }(892)^0 K^- $ production via other possible mechanisms and estimate its influence on the  $ \pi^-  K^+$ and  $ K^{+} K^- $  systems. For example, in terms of hadrons it might be thought as sequential two-body decays: the first one being $B^- \to  K^{\ast }(892)^0 K^-  $, and the second one the $K^{\ast }(892)^0 \to K^{+} \pi^- $, as shown in Fig.~\ref{fig4}.

The amplitude associated to the reaction in Fig.~\ref{fig4} can be calculated by making use of effective SU$(3)$ invariant structures of the type $  ([ P , \partial_{\mu} P] V^{\mu} ) $, where $V^{\mu}$ denotes the vector meson field and $P$ the pseudoscalar meson field. After proceeding in a usual way, this amplitude can then be written as 
\begin{eqnarray}
 t^{\prime} & = & \alpha \left[-( p_{B^-} +  p_{K^-} )\cdot ( p_{K^+} - p_{\pi^-} ) + \frac{q \cdot ( p_{B^-} +  p_{K^-} ) \, q \cdot ( p_{K^+} - p_{\pi^-} )}{m^{2}_{K^*}} \right] \frac{1}{q^2 - m^2_{K^*} + i m_{K^*} \Gamma_{K^*} } ,	
    \label{eq10}
\end{eqnarray}
where $ q^{\mu}=( p_{B^-} -  p_{K^-} )^{\mu} =( p_{K^+} + p_{\pi^-} )^{\mu} $ and $ \Gamma_{K^*}$ are the momentum and the width of the intermediate $ K^* $ meson, respectively; and $ \alpha $ is a parameter to be fixed from the data. It is convenient to define the variables 
\begin{eqnarray}
 s_{12}& = & (  p_{\pi^-} + p_{K^+} )^2 = m_{\pi}^2 + m_{K}^2 + 2 p_{\pi^-} \cdot p_{K^+}   , \nonumber \\
 s_{23}& = & (  p_{K^+} + p_{K^-} )^2 = 2 m_{K}^2 + 2 p_{K^+} \cdot p_{K^-}   , \nonumber \\
 s_{13}& = & (  p_{\pi^-} + p_{K^-} )^2 = m_{\pi}^2 + m_{K}^2 + 2 p_{\pi^-} \cdot p_{K^-}   .
    \label{eq11}
\end{eqnarray}
Then, making use of the relations above and $ s_{12}+ s_{23}+ s_{13}= m_{B}^2 +2 m_{K}^2 + m_{\pi}^2 $, we obtain
\begin{eqnarray}
( p_{B^-} +  p_{K^-} )\cdot ( p_{K^+} - p_{\pi^-} )  & = & ( p_{K^+} + p_{\pi^-} +  2 p_{K^-})  \cdot ( p_{K^+} - p_{\pi^-} )  \nonumber \\
& = & s_{12} + 2 s_{23} - m_{B}^2 - 2 m_{K}^2 - m_{\pi}^2 ; \nonumber \\
q \cdot ( p_{B^-} + p_{K^-} ) & = & m_{B}^2 -  m_{K}^2 ;  \nonumber \\
q \cdot ( p_{K^+} - p_{\pi^-} ) & = & m_{K}^2 - m_{\pi}^2 . 
    \label{eq11bis}
 \end{eqnarray}
Hence, with these last expressions, Eq.~(\ref{eq10}) can be rewritten as 
\begin{eqnarray}
 t^{\prime} & = & \alpha \left[ m_{B}^2 + 2 m_{K}^2 + m_{\pi}^2 - s_{12} - 2 s_{23}  + \frac{ ( m_{B}^2 -  m_{K}^2 ) ( m_{K}^2 - m_{\pi}^2 )}{m^{2}_{K^*}} \right] \frac{1}{s_{12} - m^2_{K^*} - i m_{K^*} \Gamma_{K^*} } .
    \label{eq12}
\end{eqnarray}

The amplitude $ t^{\prime} $ in Eq.~(\ref{eq12}) has two variables,  $s_{12}, s_{23}$. In order to obtain its mass distribution with respect to one of these variables, we employ the master formula of the PDG~\cite{Workman:2022ynf}, 
\begin{eqnarray}
  \frac{d\Gamma}{ds_{12} ds_{23}}= \frac{1}{(2\pi)^3}\frac{1}{32 m^3_{B}}  | t^{\prime} |^2 . 
      \label{eq13}
\end{eqnarray}

We remark that the experimental data from LHCb in~\cite{LHCb:2019xmb} are related to $ \frac{d\Gamma}{ds_{12} }$ and $ \frac{d\Gamma}{ds_{23}} $. In this sense, to calculate the mass distribution with respect to a specific variable, one should fix this quantity and then integrate over the other variable. Strictly speaking, one can evaluate the mass distribution with respect to the invariant mass of $\pi^-K^+$, i.e. $  \frac{d\Gamma}{ds_{12}} $, from the integration 
\begin{eqnarray}
  \frac{d\Gamma}{dM_{\rm inv}^2(\pi^-K^+)} \equiv \frac{d\Gamma}{ds_{12}}= \int _{s_{23,min}}^{s_{23,max}} ds_{23} \,  \frac{d\Gamma}{ds_{12} ds_{23}} ,
      \label{eq14}
\end{eqnarray}
where the limits of $ s_{23} $ are~\cite{Workman:2022ynf}
\begin{eqnarray}
 s_{23,min} & = & (E_{K^+}^{\ast }+E_{K^-}^{\ast})^2 - \left( \sqrt{E_{K^+}^{\ast 2}-m_{K}^2}+\sqrt{E_{K^-}^{\ast 2}-m_{K}^2} \right)^2 , \nonumber \\
 s_{23,max} & = & (E_{K^+}^{\ast }+E_{K^-}^{\ast})^2 - \left( \sqrt{E_{K^+}^{\ast 2}-m_{K}^2}-\sqrt{E_{K^-}^{\ast 2}-m_{K}^2} \right)^2 ,
      \label{eq15}
\end{eqnarray}
with $ E_{K^+}^{\ast } $ and $ E_{K^-}^{\ast } $ being defined as,
\begin{eqnarray}
E_{K^+}^{\ast } & = & \frac{1}{2\sqrt{s_{12}}}\left( s_{12} - m_{\pi}^2 + m_{K}^2 \right) , \nonumber \\
E_{K^-}^{\ast } & = & \frac{1}{2\sqrt{s_{12}}}\left( m_{B}^2 - s_{12} - m_{K}^2 \right).  
      \label{eq16}
\end{eqnarray}
The variable $ s_{12} $  is defined in the range $s_{12} \in [(m_{\pi} + m_{K})^2, (m_{B} - m_{K})^2]$ .

Additionally, the mass distribution with respect to the invariant mass of $K^+ K^-$,  $  \frac{d\Gamma}{ds_{23}} $, can also be obtained by applying the same procedure reported above between Eqs.~(\ref{eq14}) and~(\ref{eq16}) with the appropriate change of the quantities:
\begin{eqnarray}
  \frac{d\Gamma}{dM_{\rm inv}^2(K^+ K^-)} \equiv \frac{d\Gamma}{ds_{23}}= \int _{s_{12,min}}^{s_{12,max}} ds_{12} \,  \frac{d\Gamma}{ds_{12} ds_{23}} ,
      \label{eq17}
\end{eqnarray}
where 
\begin{eqnarray}
 s_{12,min} & = & (E_{K^+}^{\prime \ast}+E_{\pi^-}^{\prime \ast })^2 - \left( \sqrt{ E_{K^+}^{\prime\ast 2}-m_{K}^2} + \sqrt{E_{\pi^-}^{\prime \ast 2}-m_{\pi}^2} \right)^2 , \nonumber \\
 s_{12,max} & = & (E_{K^+}^{\prime \ast}+E_{\pi^-}^{\prime \ast })^2 - \left( \sqrt{ E_{K^+}^{\prime\ast 2}-m_{K}^2} - \sqrt{E_{\pi^-}^{\prime \ast 2}-m_{\pi}^2} \right)^2 , 
      \label{eq18}
\end{eqnarray}
with 
\begin{eqnarray}
E_{K^+}^{\prime \ast } & = & \frac{1}{2}\sqrt{s_{23}} , \nonumber \\
E_{\pi^-}^{\prime \ast } & = & \frac{1}{2\sqrt{s_{23}}}\left( m_{B}^2 - s_{23} - m_{\pi}^2 \right).  
      \label{eq19}
\end{eqnarray}
In this case $ s_{23} $  is defined in the range $s_{23} \in [ 4 m_{K}^2, (m_{B} - m_{\pi})^2]$ .

Hence, the final expression of the mass distribution with respect to the invariant mass of $K^+ K^-$, including both contributions coming from Eqs.~(\ref{eq6}) and ~(\ref{eq12}), is given by Eq.~(\ref{eq17}) but replacing the amplitude $ t^{\prime } $ in Eq~.~(\ref{eq13}) by $ \tilde{t} $, where 
\begin{eqnarray}
\tilde{t}_{K^{+}K^{-}} = t_{K^{+}K^{-}} + t^{\prime } .  
      \label{eq20}
\end{eqnarray}

\section{Results}            \label{sec:3}       

\subsection{External emission mechanism}\label{sec:3.1}  

\begin{figure}
\centering
\includegraphics[width=0.4\columnwidth]{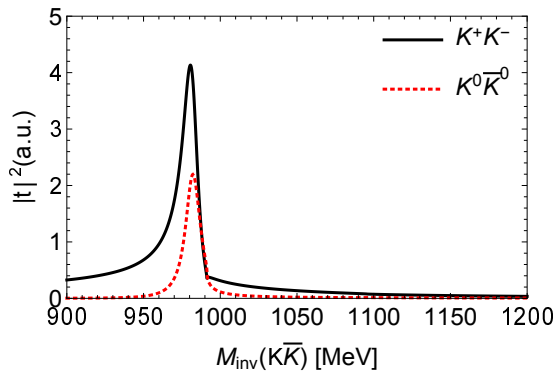}
\caption{Squared modulus of the transition matrices $t$ given by Eq.~(\ref{tranmatrix}), in arbitrary units, as functions of invariant mass of $ K^{+}K^{-} $ of  $ K^{0}\bar K^{0} $.}
\label{fig5}
\end{figure}

In Fig.~\ref{fig5} we show the squared modulus of the transition matrices $t$ given by Eq.~(\ref{tranmatrix}) as functions of invariant mass of $ K^{+}K^{-} $ and  $ K^{0}\bar K^{0} $, in arbitrary units. 
As expected, these amplitudes have peaks near the invariant mass of $ 980 \, \rm MeV $,  due to the effects of the $ I=0 $ and  $ I=1 $ contributions in the rescattering contributions (i.e. in the unitary coupled-channel amplitudes $ T $) which generate the states $f_0(980)$ and $a_0(980)$, respectively.

\begin{figure}
\centering
\includegraphics[width=0.4\columnwidth]{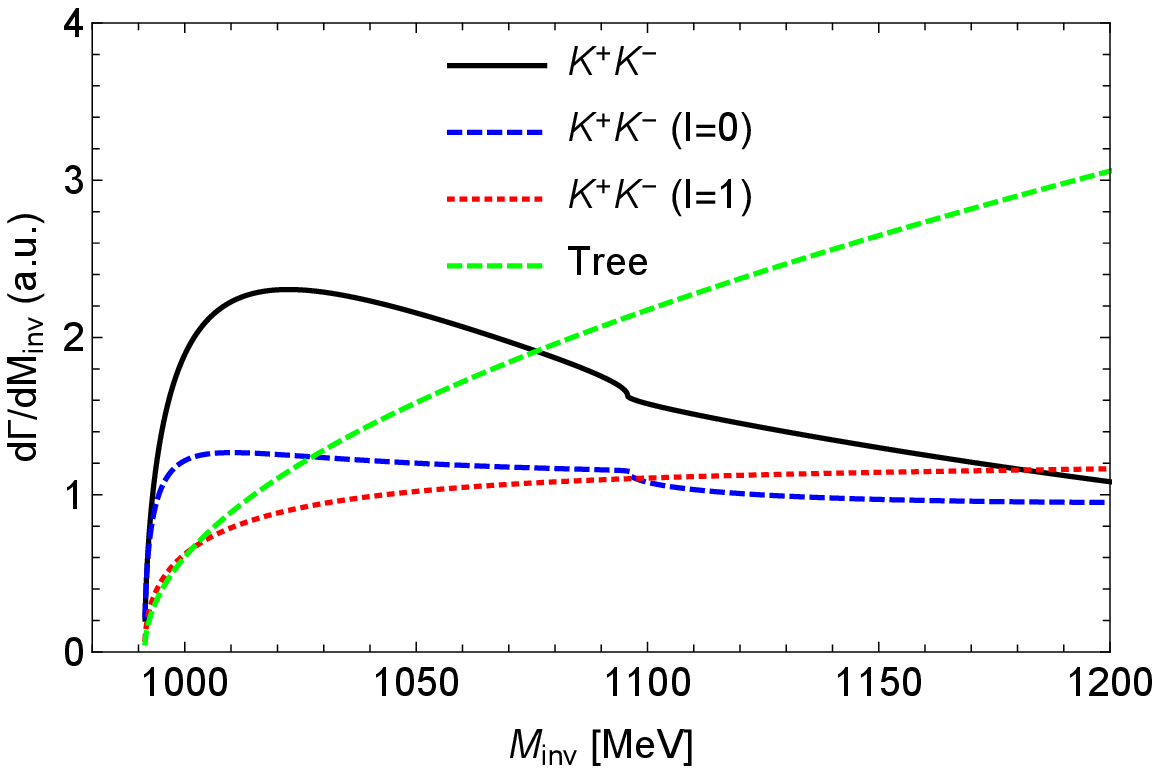}
\includegraphics[width=0.4\columnwidth]{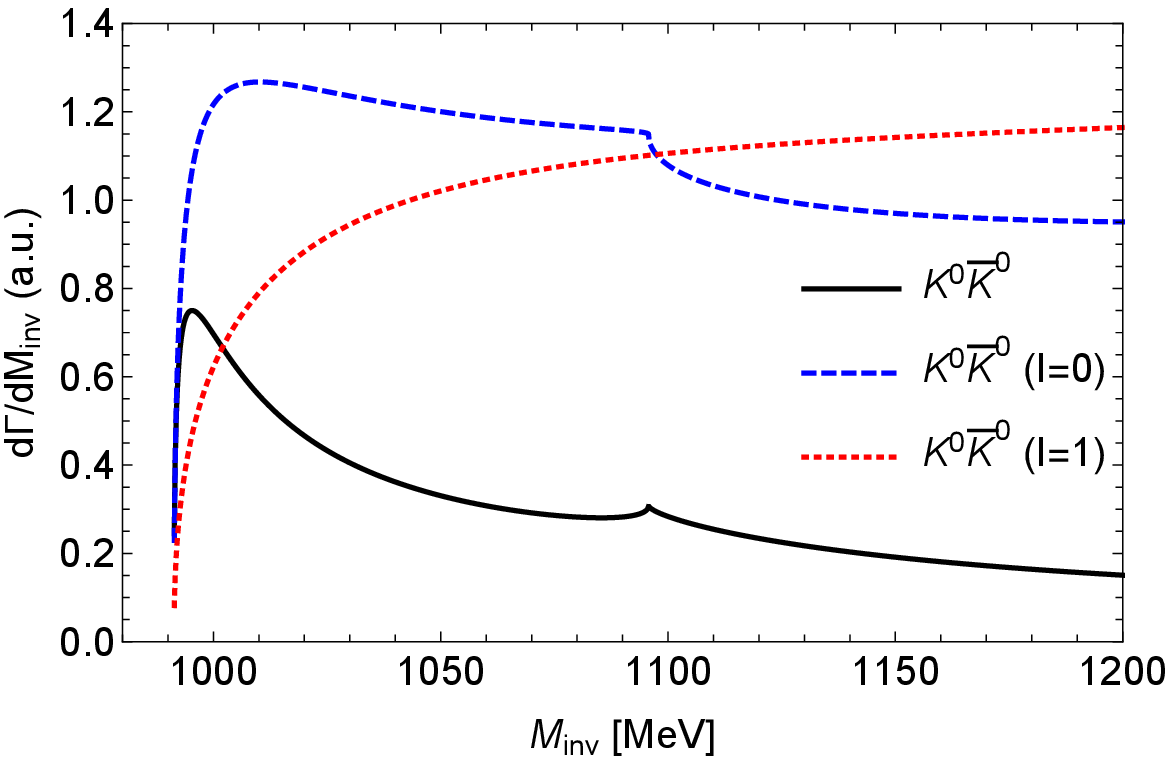}
\caption{Mass distributions $ d \Gamma_j / M_{\rm inv} $ of Eq.~(\ref{eq7}) for $ j = K^{+}K^{-} $ and  $ K^{0}\bar K^{0} $ in arbitrary units.}
\label{fig6}
\end{figure}

In the following, we present in Fig.~\ref{fig6} the plots of the mass distributions $ d \Gamma_j / d M_{inv} $ of Eq.~(\ref{eq7}) for $ K^{+}K^{-} $ and  $ K^{0}\bar K^{0} $ production in arbitrary units. To have a better understanding of the influence coming from $ I=0 $ and  $ I=1 $, they are also plotted separately. For the $ K^{+}K^{-} $ production, the tree-level contribution is also plotted. In this case of $ K^{+}K^{-} $  we remark the constructive character of the interference between the $ I=0 $ and  $ I=1 $ terms. Most importantly,  the $I=0 $ component is dominant closer to the threshold. On the other hand, the destructive interference between the $ I=0 $ and  $ I=1 $ components is manifested for the $ K^{0}\bar K^{0} $ production.

\subsection{Contribution of the channel $ K^{\ast }(892)^0 K^- $}\label{sec:3.2}  

\begin{figure}
\centering
\includegraphics[width=0.4\columnwidth]{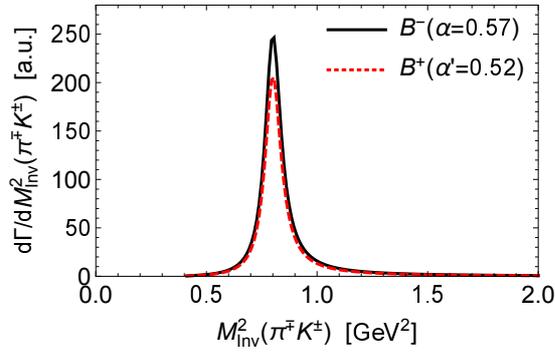}
\caption{Mass distribution $ d \Gamma / d M_{\rm inv}^2( \pi^{\mp} K^{\pm} ) $ defined in Eq.~(\ref{eq14}), in arbitrary units.}
\label{fig7}
\end{figure}

In order to estimate the impact of the reaction $ B^- \to   K^{\ast }(892)^0 K^- \to K^{+} \pi^-  K^- $, displayed in Fig.~\ref{fig4}, to the mass distribution of the $ K^{+}K^{-} $ production, we must fix the parameter $ \alpha $ of the amplitude $ t^{\prime} $ in Eq.~(\ref{eq12}). To do this, the following strategy is adopted: we employ the mass distribution $ d \Gamma / d M_{\rm inv}^2(\pi^-K^+) $ defined in Eq.~(\ref{eq14}), and determine the value of $ \alpha $ which allows us to reproduce the magnitude of the peak seen in the Fig.~2 of Ref.~\cite{LHCb:2019xmb}, which is associated to the $ K^{\ast}(892)^0 $ in $M_{\rm inv}^2(\pi^-K^+)$ for the $B^-$ decay. However, since in the mentioned figure the $B^-$ and $B^+$ data are presented separately with some differences, we apply the same method in the analysis of the peak seen in $M_{\rm inv}^2(\pi^+K^-)$ for the $B^+$ data, and set the corresponding parameter to this case, denoted here by $ \alpha^{\prime} $. The results are shown in Fig.~\ref{fig7}, already with the values of the parameters that give a reasonable fit to the LHCb data: $ \alpha = 0.52 $ and $ \alpha^{\prime} =0.57 $.

\begin{figure}
\centering
\includegraphics[width=0.4\columnwidth]{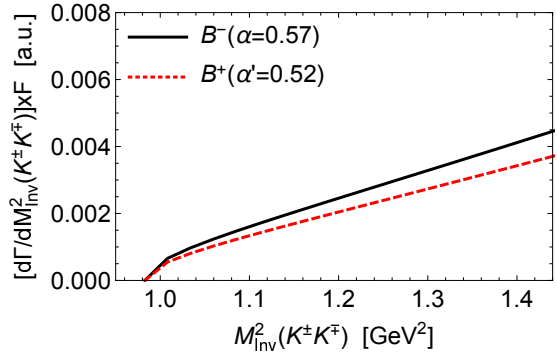}
\caption{Mass distribution $ d \Gamma / d M_{\rm inv}^2( K^{\pm}K^{\mp} ) \times F $ given by Eq.~(\ref{eq17}), in arbitrary units.}
\label{fig8}
\end{figure}

Next, using these values of $ \alpha $ and $ \alpha^{\prime} $, we evaluate the mass distribution $ d \Gamma / d M_{\rm inv}^2( K^{\pm}K^{\mp}) $ defined in Eq.~(\ref{eq17}). It is shown in Fig.~\ref{fig8}. Note, however, that although the outputs are presented in arbitrary units, in order to make a fair comparison with Fig.~3 of Ref.~\cite{LHCb:2019xmb}, we have multiplied this mass distribution by the factor $F=(0.0675/0.325)$, since the data displayed in Figs.~2 and ~3 of~\cite{LHCb:2019xmb} have different bins. 
As we can see, comparing the results of Fig.~\ref{fig8} with those of Fig.~3 of Ref.~\cite{LHCb:2019xmb}, with magnitudes $40-80$ in the same units of Fig.~\ref{fig8}, we conclude that the contribution of this $B^-,B^+$ decay channel has a negligible contribution to the $ K^{+}K^{-} (K^{-}K^{+}) $ mass distributions in the range $  M_{\rm inv}^2( K^{\pm}K^{\mp}) \in [1.0,1.44] \, \rm GeV^2 $. 
In other words, the background of the $ K^{+}K^{-} (K^{-}K^{+}) $ system coming from the  $ K^{\ast }(892)^0 K^{\mp} $ production is completely negligible in that region. Another feature is that in the range of energy of our interest (near the threshold up to $1.2 \, \rm GeV$), the difference of magnitude between the distribution for $B^-$ and $B^+$ in this decay channel is not significant.

\subsection{Distribution of $ M_{\rm inv}^2( K^{\pm}K^{\mp} ) $}\label{sec:3.3}  

\begin{figure}
\centering
\includegraphics[width=0.4\columnwidth]{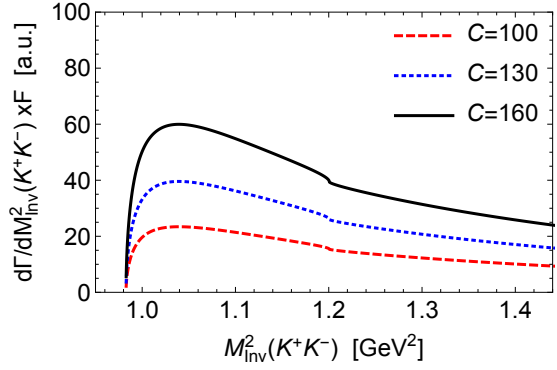}
\caption{Mass distribution $ d \Gamma / d M_{\rm inv}^2( K^{\pm}K^{\mp} ) $ given by Eq.~(\ref{eq17}) for the amplitude $ \tilde{t}_{K^{+}K^{-}} $ in Eq.~(\ref{eq20}), in arbitrary units, taking different values of parameter $C$.}
\label{fig9}
\end{figure}

\begin{figure}
\centering
\includegraphics[width=0.4\columnwidth]{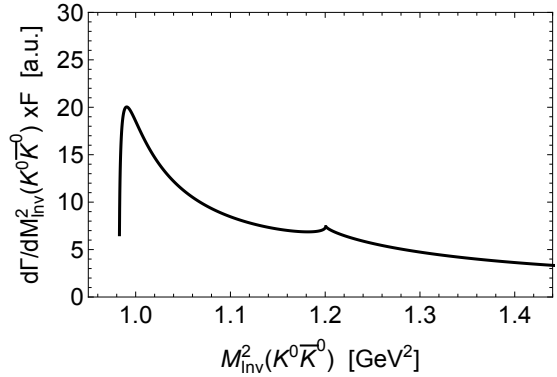}
\caption{Mass distribution $ d \Gamma / d M_{\rm inv}^2( K^0 \bar K^0 ) $ for the amplitude $ \tilde{t}_{K^0 \bar K^0} $ in Eq.~(\ref{eq20}), in arbitrary units, taking $C =160$.}
\label{fig10}
\end{figure}

Once the scheme above is already completed, we can estimate the final distribution of $ M_{\rm inv}^2( K^{\pm}K^{\mp} ) $ for the  $B^- \to  \pi ^{\mp} K^{\pm}K^{\mp}  $ reaction. 
We make use of the amplitude $ \tilde{t}_{K^{+}K^{-}} $ defined in Eq.~(\ref{eq20}) in the mass distribution $ d \Gamma / d M_{\rm inv}^2( K^{\pm}K^{\mp} ) $ given by Eq.~(\ref{eq17}), and adjust the parameter $ C $ in order to reproduce the data reported in Fig.~3 of Ref.~\cite{LHCb:2019xmb}. But keeping in mind that the difference of magnitude between the mass distributions for $B^-$ and $B^+$ generated by the $ K^{\ast }(892)^0 K^{\mp} $ production is small, here we calculate the mass distribution for the average for $B^-$ and $B^+$  and take the averaged $B^-$ and $B^+$ data of~\cite{LHCb:2019xmb} as a guide~\footnote{The magnitude of $ d \Gamma / d M_{\rm inv}^2 $ for $B^-$ and $B^+$ production are different in Fig.~3 of Ref.~\cite{LHCb:2019xmb} due to $CP$ violation. In our formalism we do not have $CP$ violation and thus, a proper comparison of our results should be made with the average of the two distributions.}. 

So, in Fig.~\ref{fig9} we plot the mass distribution $ d \Gamma / d M_{\rm inv}^2( K^{\pm}K^{\mp} ) $ for the amplitude $ \tilde{t}_{K^{+}K^{-}} $ in arbitrary units, taking different values of the parameter $C$, our normalization factor in Eq.~(\ref{tranmatrix}). 
These outputs indicate that the case with $C=160$ gives a reasonable concordance with  the experimental results, since the maximum strength of the mass distribution is similar to the averaged $B^-$ and $B^+$ data of~\cite{LHCb:2019xmb}. It is also interesting to observe that the fall down of the mass distribution is similar to the one for $ B^+ \to K^{-} K^{+} \pi^{+}$ distribution in Fig.~3 of Ref.~\cite{LHCb:2019xmb} which has more statistical significance than its complex conjugate reaction. 


We must emphasize the most important feature of these results: they show clearly the relevance of the effects due to the $f_0(980)$ and $a_0(980)$ states in the $ M_{\rm inv}^2( K^{\pm}K^{\mp} ) $ distribution in the studied decay, in particular those coming from the  $f_0(980)$ which dominates in the region of low $ K^{\pm}K^{\mp}$ invariant mass. 
The arguments used to show that the decay channel $ B^- \to  K^{\ast }(892)^0 K^- $ has a negligible contribution in the region of small $  K^{+}K^{-} $ invariant masses can equally be applied to other decay modes like the $  \pi^- K^{\ast }_0(700), \pi^- K^{\ast }_0(1430) $. The $B^- \to \pi ^- u \bar u $ in Fig.~\ref{fig1} has no overlap with $ \pi^- \phi  $ since $ \phi = s \bar s $; and the decay  $ B^- \to \pi ^- f_2(1270) $ should be highly suppressed since the $f_2(1270)$ decays to $ \pi \pi $ with a very small fraction to  $ K \bar K $~\cite{gengvec}. 

By using the value of $C=160$ which gives a fair reproduction of the data, we show in Fig.~\ref{fig10} the results for  $K^0 \bar K^0 $ production in $B^- \to \pi ^- K^0 \bar K^0 $ decay. This is a prediction based on our picture with the normalization of  Fig.~\ref{fig9}, where we see a smaller strength than for the $  K^{+}K^{-} $ production and a sharp peak very close to threshold coming from the interference of the  $f_0(980)$ and $a_0(980)$ resonances. 

With all these arguments we conclude that the mechanism responsible for the  $  K^{+}K^{-} $  mass distribution close to the $ K \bar K $ threshold in $B^- \to \pi ^- K^+ K^- $ decay is due to the production of the  $f_0(980)$ and $a_0(980)$ resonances. Our results in Fig.~\ref{fig6} indicate that in the cases of $  K^{+}K^{-} $ and $K^0 \bar K^0 $ production the $f_0(980)$ is more important than the $a_0(980)$. We have also shown that the pattern of 
 $  K^{+}K^{-} $ and  $ K^{0}\bar K^{0} $ production are very different, having a constructive interference of the $I=0$ and $I=1$ components for $  K^{+}K^{-} $ production and a destructive interference for the case of $ K^{0}\bar K^{0} $. The measurement of the $B^- \to \pi ^- K^0 \bar K^0 $ decay would, thus, be an important complement to show the relevance of the  $f_0(980)$ and $a_0(980)$ production in these decays and its relationship to the dynamical origin of these resonances.

\section{Concluding remarks}            \label{sec:5}       

In this work we have analyzed the role of the $f_0(980)$ and $a_0(980)$ resonances in the low $ K ^{+} K^{-} $ invariant-mass region of the $B^- \to \pi ^- K^+ K^- $ and $B^- \to \pi ^- K^0 \bar K^0 $ reactions. We have made use of the chiral unitary $\rm SU(3)$ formalism, in which these two resonances are dynamically generated from the unitary pseudocalar-pseudoscalar coupled-channel approach. Then,  the amplitudes and the mass distributions with respect to the  $  K^{+}K^{-} $ and  $ K^{0}\bar K^{0} $ invariant-masses have been calculated, with the contributions coming from the $I=0$ and $I=1$ components being explicitly evaluated. For completeness, the contribution of the $ K^{\ast }(892)^0 K^- $ production and its influence on the  $ \pi^-  K^+$ and  $ K^{+} K^- $  systems have also been computed, not presenting a relevant contribution in the region of small $ K \bar K $ invariant mass. Finally, the distributions of $ M_{\rm inv}^2( K^{\pm}K^{\mp} ) $ for the  $B^{\mp} \to  \pi ^{\mp} K^{\pm}K^{\mp}  $ reaction have been estimated and compared with the LHCb data in~\cite{LHCb:2019xmb}. Our findings indicate that the low $ K ^{+} K^{-} $ invariant-mass region has a leading contribution coming from the $I=0$ component through the $f_0(980)$ excitation.

We have discussed the contribution of other channels in the region of low $ K \bar K $ mass distributions, concluding that the formation of the $f_0(980)$ and $a_0(980)$ resonances in the $B^- \to \pi ^- f_0(980) $ and $B^- \to \pi ^- a_0(980) $ decays are largely responsible for the strength of the $ K \bar K $ mass distribution in that region. Within the same framework, we have also evaluated the $ K^{0}\bar K^{0} $ distribution in the $B^- \to \pi ^- K^0 \bar K^0 $ decay and found a smaller strength than for $B^- \to \pi ^- K^0 \bar K^0 $ and a shape quite different to the latter one. This is a consequence of a constructive or destructive interference of the resonances in the $  K^{+}K^{-} $ and  $ K^{0}\bar K^{0} $ production. 
We believe that these results deserve to be tested and evaluated in the future experimental works.

\section*{Acknowledgments}

The work of L.M.A. was partly supported by the Brazilian agencies CNPq (Grant Numbers 309950/2020-1, 400215/2022-5, 200567/2022-5), FAPESB (Grant Number INT0007/2016) and CNPq/FAPERJ under the Project INCT-Física Nuclear e Aplicações (Contract No. 464898/2014-5). The work of N. I. was partly supported by JSPS KAKENHI Grant Numbers JP19K14709 and JP21KK0244.
This work is also partly supported by the Spanish Ministerio de
Economia y Competitividad (MINECO) and European FEDER funds under Contracts No. FIS2017-84038-C2-1-P
B, PID2020-112777GB-I00, and by Generalitat Valenciana under contract PROMETEO/2020/023. This project has
received funding from the European Union Horizon 2020 research and innovation programme under the program
H2020-INFRAIA-2018-1, grant agreement No. 824093 of the STRONG-2020 project.


\end{document}